\input{aipcheck}

\documentclass[
    ,final            
  ]
  {aipproc}

\layoutstyle{8x11double}

\begin{document}

\title{Central engines of Gamma Ray Bursts. Magnetic mechanism in the collapsar
model.}


\author{Maxim V.~Barkov}{
  address={Department of Applied Mathematics, The University of Leeds,
Leeds, LS2 9GT,UK}
 ,altaddress={Space Research Institute, 84/32 Profsoyuznaya Street, Moscow
117997, Russia} 
}

\author{Serguei S.~Komissarov}{
  address={Department of Applied Mathematics, The University of Leeds,
Leeds, LS2 9GT,UK}
}

\begin{abstract}
In this study we explore the magnetic mechanism of hypernovae  
and relativistic jets of long duration gamma ray bursts within the collapsar 
scenario.  This is an extension of our earlier work\cite{bk08a}. 
We track the collapse of massive rotating stars onto a rotating central black 
hole using axisymmetric general relativistic magnetohydrodynamic code that 
utilizes a realistic equation of state and 
takes into account the cooling associated with emission of neutrinos
and the energy losses due to dissociation of nuclei. The neutrino heating
is not included.  We describe solutions with different 
black hole rotation, mass accretion rate, and strength of progenitor's 
magnetic field. Some of them exhibits strong explosions driven by 
Poynting-dominated jets with power up to $12\times10^{51}\,\mbox{erg s}^{-1}$. 
These jets originate from the black hole and powered via
the Blandford-Znajek mechanism. A provisional criterion for explosion is derived. 
A number of simulation movies can be downloaded from
http://www.maths.leeds.ac.uk/~serguei/research/movies/anim.html  
\end{abstract}

\maketitle


\section{Introduction}
\label{introduction}

The most popular model of central engines of long-soft Gamma Ray Burst 
(GRB) is based on the ``failed supernova''
scenario of stellar collapse, or ``collapsar'', where the core of a
rapidly rotating progenitor star forms a black hole (BH) surrounded by  
an accretion disk \cite{W93}.  A similar configuration can be
produced via inspiraling of a BH or a neutron star into the companion star
during the common envelope phase of close binary \cite{FW98,ZF01}.  The
gravitational energy released in the disk can be very large, more than
sufficient to stop the collapse of outer layers and drive GRB outflows,
presumably in the polar direction where density is much lower \cite{MW99}.
GRB jets can be powered via heating due to annihilation of disk neutrinos \cite{MW99} 
and via magnetic braking \cite{BP82,UM06,PMAB03}. 

The high angular momentum assumed for the stellar core in the collapsar model
also implies high rotation rates of the central black hole and the
possibility of powering the GRB jets via the Blandford-Znajek mechanism
\cite{BZ77}. The rotational energy of black hole with mass $M_{h}=2M_{\odot}$ 
and spin $a=0.9$ is enormous, $E_{rot} \simeq5\times10^{53}$erg, which is
fifty times higher than the rotational energy of a millisecond pulsar
\cite{KB07} and well above what is needed to explain GRBs 
and associated hypernovae. Applying the Blandford-Znajek formula to the
collapsar problem we obtain the following estimate of power
\begin{equation} \dot{E}_{BZ}=3.6\times10^{50} f(a) \Psi_{27}^2
M^{-2}_{2} \,\mbox{erg}\, \mbox{s}^{-1},
\label{ebz}
\end{equation} 
where $M_{2}=M_{h}/2M_\odot$, $\Psi_{27}=\Psi/10^{27}\mbox{G}\,\mbox{cm}^2$ is 
the magnetic flux penetrating BH, 
$f(a)=a^2/(1+\sqrt{1-a^2})^2$, and it is assumed that BH magnetosphere rotates with 
angular velocity $\Omega=0.5\Omega_h$, thus delivering maximum power. 
These estimates show that the magnetic
braking of BH  could fully account for the energetics of GRBs. However, 
the result $\Omega=0.5\Omega_h$ is obtained for force-free
magnetospheres where one can ignore the inertia of matter and it is not clear
when and how such conditions can develop in the collapsar model. For high
mass-loading in the magnetosphere one could expect the magnetospheric rotation 
to be dictated by the angular momentum of matter rather then by the rotation of BH
itself. Careful GRMHD simulations are needed in order to resolve this issue.
Previous studies, with simplified setting and micro physics, have already
demonstrated 
that this problem can be addressed with modern computational tools (e.g.\cite{M06}) 
and invite to explore progressively more realistic models. Here we present 
numerical solutions for models with realistic EOS and effects of neutrino cooling.     
The neutrino transport, however, is rather complicated, particularly in the curved 
space of BH, and for this reason we have not included the heating due to neutrino 
annihilation. In reality, both the neutrino and magnetic mechanisms could operate 
hand in hand and we plan to include neutrino heating in future studies. 
The results for one particular numerical model have been presented in \cite{bk08a};  
here we outline a wider study. The effects of neutrino cooling on disc
accretion 
have been described in \cite{bk08c}.

\section{Simulation Setup}
\label{setup}

Since our code can deal only with time-independent space-time we are forced to
start from the point where the central BH of mass $M_{h}=3M_{\odot}$ has
already been formed inside the collapsing star.  The rotation rate of BH is
rather uncertain and depends on the angular momentum distribution in the
progenitor. Here we use either the rather optimistic value of $a=0.9$ or the
very pessimistic value of $a=0.0$. The latter was introduced to fully
eliminate the Blandford-Znajek effect and to see if the magnetic braking of
accretion disk alone could make the star to explode.

The main details of our numerical method and various test simulations are
described in \cite{K04b}. The only really new feature
here is the introduction of HLL-solver which is activated when our linear
Riemann solver fails, this occurs in regions of relativistically high
magnetization. We have also found that total switch
to the HLL-solver significantly degrades numerical solutions via increasing
numerical diffusion and dissipation (see also \cite{MB06}).

The gravitational attraction of BH is introduced via Kerr metric in
Kerr-Schild coordinates, $\{t,\phi,r,\theta\}$.  The two-dimensional
computational domain is $(r_0<r<r_1)\times(0<\theta<\pi)$, where $r_0 =
(1+0.5\sqrt{1-a^2}) \; r_g$ and $r_1=5700 \;r_g = 25000$ km. 

Notice that the inner boundary is inside of the outer event horizon -
we can do this because the horizon coordinate singularity is eradicated in the
Kerr-Schild coordinates.  The total mass within the domain is small compared
to the mass of BH ( only about $7\%$) that allows us to ignore its self-gravity.  The
grid is uniform in $\theta$ where it has 180 cells and almost uniform in
$\log(r)$ where it has 450 cells, the linear cell size being the same in both
directions.

A realistic equation of state (EOS) is introduced using the EOS code
HELM\footnote{ the code can be downloaded from the website
http://www.cococubed.com/code\_pages/eos.shtml} \cite{TS00}. 
The neutrino cooling is computed using the interpolation formulas given in 
\cite{IIN69,S87,bezchas}. The photo-disintegration of nuclei is included via
modifying EOS as in \cite{ABM05}. 
The equation for mass fraction of free nucleons is adopted from
\cite{WB92}.


The collapsing star is described by the  free-fall model of
 \cite{bethe}.  The collapse time for a core of radius $r_c=10^9$cm 
and mass 3.0$M_{\odot}$ is
$t_c=\sqrt{2}r_c/3\mbox{v}_{\mbox{ff}}\simeq 0.5$s. Since, in this study we
explore the
possibility of explosion soon after the core collapse, we start simulations 
1 second after the onset on collapse.
Because GRBs are currently associated with very massive progenitors
we set the mass parameter $C_1$ of Bethe's model (see \cite{bk08a} ) 
to either $C_1=3$ or $C_1=9$.  
These correspond to the accretion rates of $ \dot{M}
\simeq 0.166 M_{\odot} \mbox{s}^{-1}$ and $ \dot{M} \simeq 0.5 M_{\odot}
\mbox{s}^{-1}$ respectively. 
The angular momentum distribution describes a solid
body rotation within the cylindrical $r_l = 6300$km and is constant with the
value of $l_0=10^{17}\mbox{cm}^2\mbox{s}^{-1}$ further out, similar to 
that used in \cite{MW99}. These initial conditions are also
used to fix the flow variables in the ghost cells of the outer boundary.

The magnetic field distribution is that of a uniformly magnetized
sphere in vacuum, the radius of this sphere $r_1=4500$km. Inside the sphere 
the magnetic field strength is $B_0=3\times10^{10}$G, $10^{10}$G or
$3\times10^{9}$G. 

\begin{figure}
\includegraphics[width=70mm]{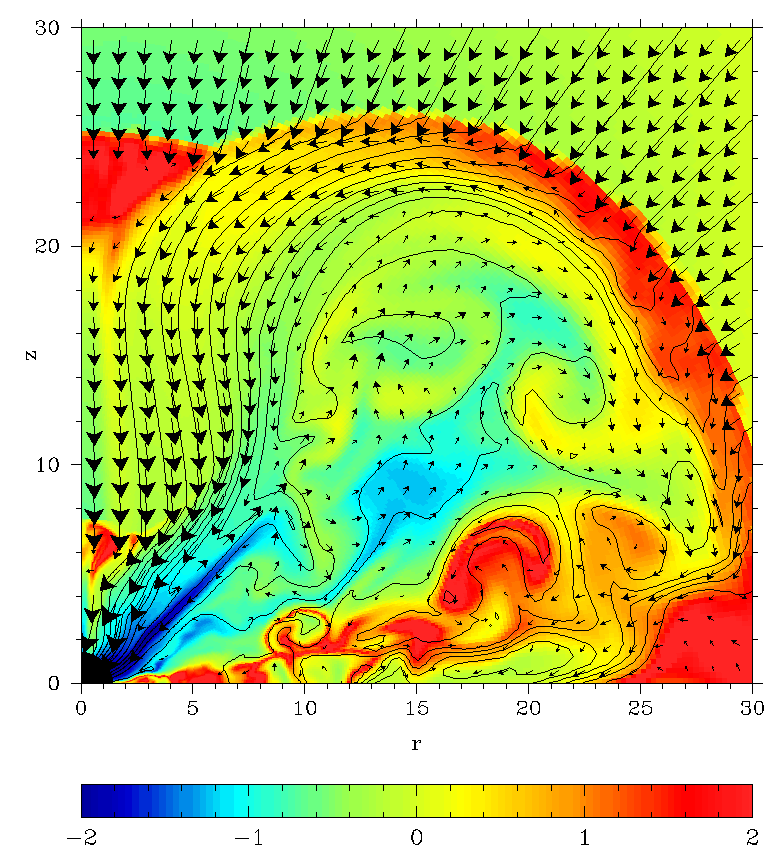}
\caption{Model $C_1=9$, $B_0=3\times 10^{10}$ G at time $t=0.24$s, 
soon before the explosion. Colour shows the ratio of gas and magnetic 
pressure, $\log_{10}(P/P_m) $, contours the magnetic field lines
and arrows the velocity field. Notice the outflow beginning to develop
in the highly magnetized ``blue stripe'' stretching at the angle of $45^o$ 
from the black hole.}
\label{c9b10}
\end{figure}

\begin{figure*}
\includegraphics[width=57mm]{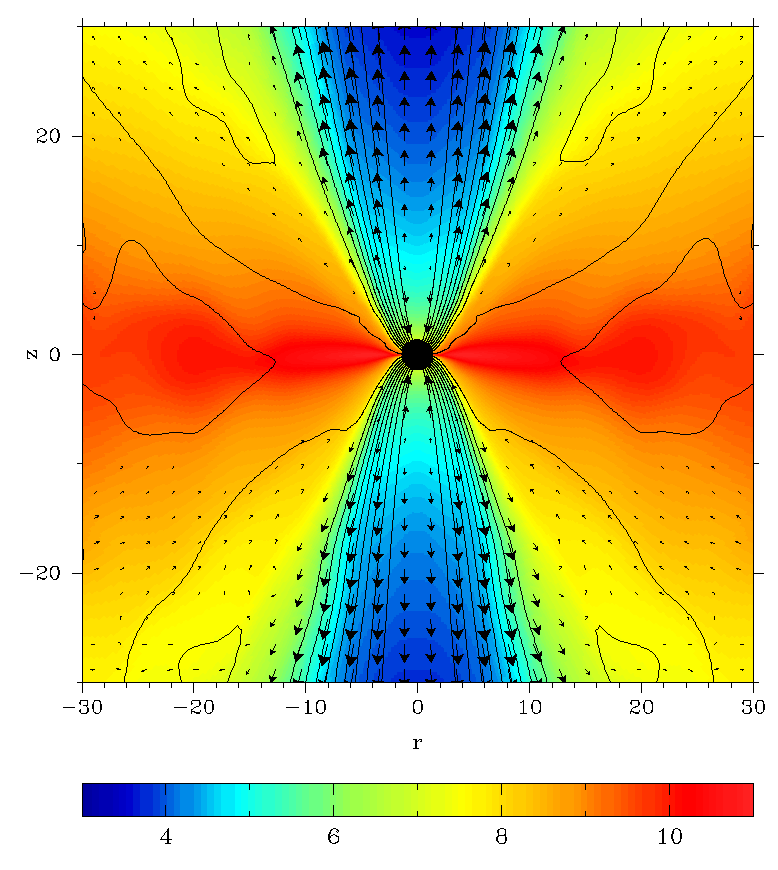}
\includegraphics[width=57mm]{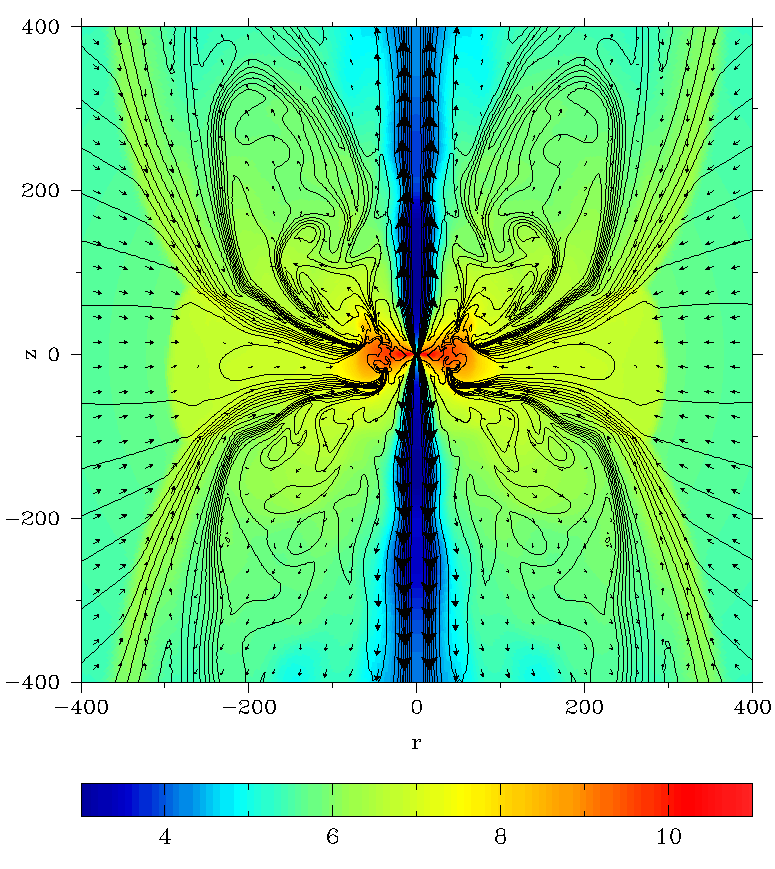}
\includegraphics[width=57mm]{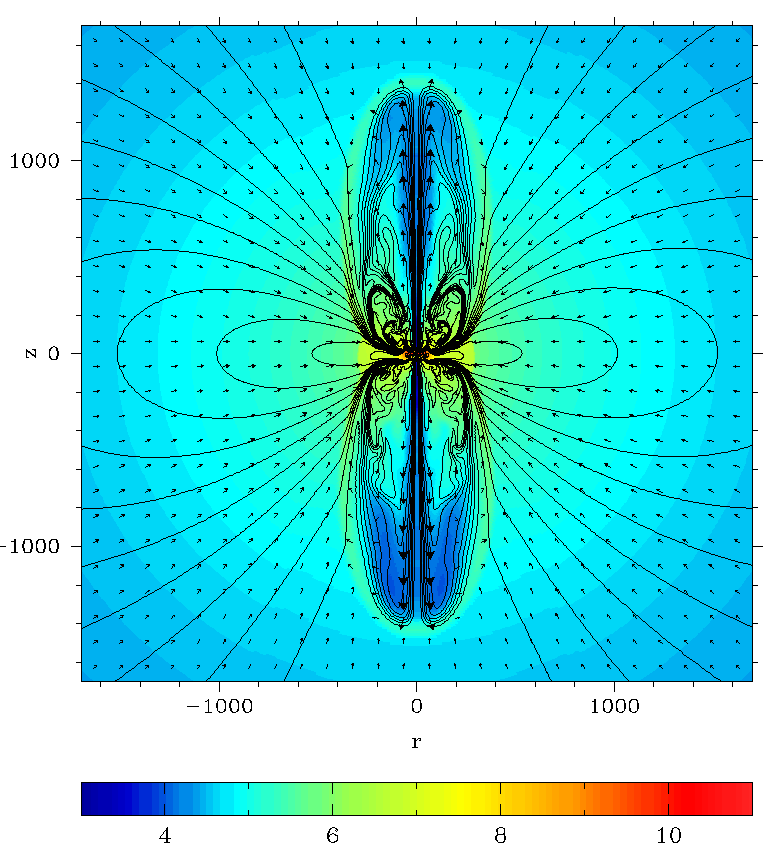}
\caption{Solution for $C_1=3$, $B_0=10^{10}$ G on different scales at $t=0.75$s.
Colour shows the baryonic rest mass density, $log_{10}\rho$ in
g/cm$^3$, contours the magnetic field lines, and arrows the
velocity field.}
\label{c3b95}
\end{figure*}


\section{Computer simulations}
\label{simulations}

At the beginning of simulations the angular momentum of 
gas near BH is less than that of the last stable orbit, $l_{lso}$, 
and as a result it falls straight into the BH. The most rapidly 
varying parameter at this stage is the magnetic flux threading the 
BH which grows linearly with time.  Later, when matter with 
$l\simeq l_{lso}$ reaches the BH it forms a thin inviscus accretion disk 
in the equatorial plane \cite{bi01}. Finally, when the angular 
momentum of gas approaching the BH in the equatorial direction well 
exceeds $l_{lso}$ the centrifugal force halts its infall and a turbulent 
accretion disk develops around the BH. At the same time a strong 
accretion shock lifts from its surface. After passing the accretion 
shock the low angular momentum plasma of polar regions keeps falling 
straight into the BH  whereas the high angular momentum plasma coming 
at intermediate directions fills the low density volume above and below 
the disk. Strong differential rotation within this bubble results 
in strong amplification of azimuthal magnetic field. 
Further evolution depends of the amount of magnetic flux accumulated by the BH. 

In the model with $C_1=9$ and $B_0=3\times10^{10}$G we observed strong explosion 
very soon after formation of the accretion disk. The BH is a key factor 
in this explosion pumping the electromagnetic energy into the bubble at the 
rate of $\simeq 12\times 10^{51}\mbox{erg}\,\mbox{s}^{-1}$. Prior to the 
explosion we see mass unloading of some magnetic field lines connecting the 
bubble with BH (see fig.\ref{c9b10}). 
This seems to be promoted by the fact plasma entering the accretion
shock along these field lines is diverted towards the equatorial plane 
instead of entering the bubble and falling into the BH.  
Similar evolution was shown by the model described in \cite{bk08a} 

In the model with $C_1=3$ and $B_0 = 10^{10} G$ we observe several pulsations 
of the accretion shock, gradual expansions followed by relatively rapid 
contractions, prior to the explosion which is significantly less energetic. 
However, once the explosion sets up it shows the same generic features 
(see fig.\ref{c3b95}). We observe two well defined
polar jets surrounded by magnetic cocoons of high pressure and low density.
The magnetic pressure of these cocoons, which have been inflated by the jets, 
exceeds by more than six orders of magnitude the magnetic pressure in 
the collapsing star.  These over-pressured cocoons drive a  
blast wave into the star. The mean propagation speed of the blast in the 
polar direction $v_{s}\simeq 0.1c$.  
In the vicinity of BH the solution shows the same key features as 
found in the previous studies of thick disks around BHs -- main disk, 
its dynamic corona, and magnetically-dominated 
funnel inside which two Pointing-dominated jets are produced 
by the BH \cite{DHK03,MG04,SST07,bk08a,ssl08,bk08c}.
Integrating over the volume of the blast wave we can measure the rate of 
energy supply in the explosion, 
$\dot{E} \simeq 0.4\times10^{51}\mbox{erg}\,\mbox{s}^{-1}$. The direct 
measurement of the energy flux across the horizon inside the funnel gives 
a very similar number,  
$\dot{E}_{BZ}\simeq 0.42\times10^{51}\mbox{erg}\,\mbox{s}^{-1}$, which is 
also in excellent agreement with eq.1. Thus, the explosion is undoubtedly 
powered by the Blandford-Znajek mechanism. 
Finally, this model seems to be marginal as any further noticeable increase 
of accretion rate or decrease of magnetic field strength prevents explosion.

Although initially the BH jets are Poynting-dominated their electromagnetic
energy is gradually converted into the energy of matter. To some extent 
this may reflect real physical processes \cite{M06,KBVK07,bk08b,tmn08} but 
numerical diffusion and resistivity are likely to play a significant 
part too. In any case, the total energy is conserved in the simulations 
and we do not expect these numerical problems to effect the dynamics of blast.

In order to investigate further the potential of magnetic breaking of  
accretion disk itself we carried out simulations with non-rotating BHs, 
thus eliminating the BZ-effect. The results show that the accretion disk is 
formed much later when almost all magnetic flux is already accreted onto 
the BH and it begins to ''swallow'' the closed magnetic field lines. 
As in other solutions, we observe lift-off of strong accretion shock, 
development of disk corona, and amplification azimuthal magnetic 
field. However, the explosion does not take place even under the most  
favourable conditions (the highest initial magnetic field and the lowest 
mass accretion rate) in our parameter space.  
This result provides an additional support to our conclusion 
on the crucial role of the BZ-mechanism.   

Our results seem to fit the following condition for magnetically driven 
explosion

\begin{equation}
 \dot{M}_{0.1}^{-0.5} B_{10} r^2_{m,9}
 \frac{a}{(1+\sqrt{1-a^2})} >0.1,
\label{se_lim}
\end{equation}
where $B_{0,10}=B_0/10^{10}$G, $r_{m,9}=r_m/10^9$cm,
$\dot{M}_{0.1}=\dot{M}/0.1 M_{\odot}\mbox{s}^{-1} $, which 
is obtained by comparing the maximum power of the BZ-mechanism 
(see eq.1) and the power of accretion flow. Further investigation is 
needed to verify this condition for intermediate values of the spin 
parameter. For $\dot{M}_{0.1}>0.5$ the power of neutrino annihilation 
becomes substantial and this may lead to a much less demanding criterion. 

\section{Discussion and Conclusions}
\label{disscusion}

Our results suggest that magnetic fields may play a crucial role in 
powering GRB jets and associated hypernovae not only in the 
magnetar but also in the collapsar model. In the latter case the main 
source of energy is the rotation of BH and it is released via 
the Blandford-Znajek mechanism. The rate of energy release seen in 
the simulations, $\dot{E}=0.2\div 16\times10^{51}\mbox{erg}\,\mbox{s}^{-1}$, 
is very high and can easily explain the energetics of hypernovae, 
$E\simeq 10^{52}$erg. The fact that the rotational energy of maximally 
rotating BH is much higher, $E_{rot}\simeq\mbox{few}\times10^{53}\mbox{erg}$, 
suggests a self-regulating process in which the rate of energy release 
significantly reduces after the first few seconds, presumably due to the lower 
mass supply into the accretion disk when the stellar collapse is reversed 
by the blast, the escape of magnetic flux from the black hole, or its annihilation 
due to the accretion of magnetic field lines of opposite polarity. This idea is 
supported by the lower observed energetics of GRB jets. With the typical 
total energy of $10^{51}$erg and duration of $10\div100$s their power 
is only $0.1\div0.01\times 10^{51}\mbox{erg}\,\mbox{s}^{-1}$ \cite{bfk03,P05}.    
Alternatively, the hypernovae blast wave could be caused mainly by the 
neutrino heating during the short initial period of high accretion rate. 
In this case much weaker magnetic field will be required to drive GRB 
jets into the cavity created by the blast.

\begin{theacknowledgments}
This research was funded by PPARC under the rolling grant
``Theoretical Astrophysics in Leeds''. The simulations were carried out 
on the St Andrews UK MHD cluster and on the White Rose Grid cluster 
Everest.
\end{theacknowledgments}



\bibliographystyle{aipproc}   

\end{document}